# Diagnostic Accuracy of Open-Source Vision-Language Models on Diverse Medical Imaging Tasks

Gustav Müller-Franzes[1], Debora Jutz[1], Jakob Nikolas Kather[2,3], Christiane Kuhl[1], Sven Nebelung[1], Daniel Truhn[1]

(1) Department of Diagnostic and Interventional Radiology, University Hospital RWTH Aachen, Aachen, Germany
(2) Else Kroener Fresenius Center for Digital Health, Technical University Dresden, Dresden, Germany
(3) Department of Medicine I, University Hospital Dresden, Dresden, Germany

## Abstract

This retrospective study evaluated five VLMs (Qwen2.5, Phi-4, Gemma3, Llama3.2, and Mistral3.1) using the MedFMC dataset. This dataset includes 22,349 images from 7,461 patients encompassing chest radiography (19 disease multi-label classifications), colon pathology (tumor detection), endoscopy (colorectal lesion identification), neonatal jaundice assessment (skin color-based treatment necessity), and retinal fundoscopy (5-point diabetic retinopathy grading). Diagnostic accuracy was compared in three experimental settings: visual input only, multimodal input, and chain-of-thought reasoning. Model accuracy was assessed against ground truth labels, with statistical comparisons using bootstrapped confidence intervals (p<.05).

Qwen2.5 achieved the highest accuracy for chest radiographs (90.4%) and endoscopy images (84.2%), significantly outperforming the other models (p<.001). In colon pathology, Qwen2.5 (69.0%) and Phi-4 (69.6%) performed comparably (p=.41), both significantly exceeding other VLMs (p<.001). Similarly, for neonatal jaundice assessment, Qwen2.5 (58.3%) and Phi-4 (58.1%) showed comparable leading accuracies (p=.93) significantly exceeding their counterparts (p<.001). All models struggled with retinal fundoscopy; Qwen2.5 and Gemma3 achieved the highest, albeit modest, accuracies at 18.6% (comparable, p=.99), significantly better than other tested models (p<.001). Unexpectedly, multimodal input reduced accuracy for some models and modalities, and chain-of-thought reasoning prompts also failed to improve accuracy.

The open-source VLMs demonstrated promising diagnostic capabilities, particularly in chest radiograph interpretation. However, performance in complex domains such as retinal fundoscopy was limited, underscoring the need for further development and domain-specific adaptation before widespread clinical application.

# Introduction

Large language models, such as ChatGPT, have demonstrated utility in a range of text-based medical applications, including report verification(1,2), simplification of complex language(3,4), and the extraction of clinically relevant information from free-text reports(5,6). However, to interpret medical images and answer questions, models capable of jointly interpreting images and accompanying text, such as vision language models (VLMs), are needed(7,8).

Recent studies have highlighted the growing role of LLMs and VLMs in clinical applications, from diagnostic dialogue systems(9) to mental health interventions(10). However, the development and deployment of state-of-the-art VLMs require extensive computational infrastructure and large-scale training datasets. Consequently, the most capable VLMs are developed by global companies, which deploy them via proprietary, cloud-based platforms. For healthcare settings, where sensitive information must remain within institutional boundaries, these platforms raise important concerns regarding patient privacy and data governance.

In light of these challenges, an ideal medical VLM must demonstrate exceptional accuracy and be resilient to the inherent variability and imperfections of real-world medical data. It should generalize well to unseen data from different hospitals or patient populations. Furthermore, an ideal medical VLM should have the ability to seamlessly handle multimodal data, as patient diagnosis is rarely based on image data alone. Given the sensitive nature of patient information, the model must also be designed to comply with strict privacy regulations.

Several companies have recently released open-source variants of their VLMs, such as Gemma3 (11), Phi-4 Multimodal (12), Llama3.2 (13), Mistral3.1(14), and Qwen2.5 (15). These distilled versions of commercial VLMs have strong performance on general-purpose vision-language benchmarks but are sufficiently lightweight to be deployed within local clinical infrastructure, including hospital servers or personal workstations. Thus, they hold significant promise for privacy-preserving, institution-controlled deployment of artificial intelligence tools in healthcare. However, none were specifically trained for medical applications, and their initial pretraining remains computationally prohibitive for most users. A systematic comparison is needed to evaluate the diagnostic utility of these open-source VLMs within clinical imaging contexts, specifically, their accuracy in interpreting medical images when used "off the shelf", without pretraining. Accordingly, the purpose of this study was to comprehensively evaluate the diagnostic accuracy of open-source VLMs in medical vision tasks to understand the extent to which these models can support diagnostic decision-making in routine clinical care.

# Materials and Methods

## Study Design

This retrospective study of the diagnostic performance of open-source VLMs in medical imaging tasks used publicly available medical imaging datasets. Model performance was evaluated under three distinct experimental settings . First, image-only inputs were used to assess the models' standalone visual capabilities, relying solely on medical images without additional contextual information. Second, multimodal inputs combining image data with structured clinical parameters were used to evaluate the models' ability to integrate and process multimodal data for clinical decision-making. Finally, chain-of-thought reasoning was employed to assess the models' ability to enhance their responses through reasoning. In this last setting, the models were prompted with tasks that required them to articulate their reasoning prior to providing an answer.

## Data Acquisition

This study utilized the publicly available MedFMC dataset, a collection designed to benchmark foundation model adaptation in medical image classification tasks(16). The dataset comprises 22,349 images from 7,461 patients across five distinct medical imaging modality subsets and corresponding clinical classification challenges. The number of patients contributing images varies by subset:

The ChestDR subset comprises 4,848 frontal chest radiographs from 4,848 patients at a single institution. These images were acquired as part of routine clinical thoracic disease screening, and the presence of 19 common thoracic diseases is labeled.

The ColonPath subset consists of 10,009 image patches from 396 patients and was derived from the pathological examination of colorectal tissue. The patches represent areas evaluated for the presence of tumor tissue, originating from surgical pathology specimens.

The Endo subset includes 3,865 images captured from 80 patients during standard colonoscopy procedures. The images are representative frames selected for the assessment of colorectal lesions, including ulcers, erosion, polyps, and tumors.

The NeoJaundice subset is a collection of 2,235 digital camera images acquired from 745 patients for the purpose of evaluating the necessity of neonatal jaundice treatment based on skin color analysis.

The Retino subset contains 1,392 fundus images from 1,392 patients evaluated for diabetic retinopathy using standard fundus photography techniques. Each image is labeled with a score for the presence of diabetic retinopathy: 0, No; 1, Mild; 2, Moderate; 3, Severe; 4, Proliferative.

## Models

All models evaluated in this study were accessed via the Hugging Face platform, ensuring a standardized and reproducible deployment environment. For each model, the "instruct" version was selected to align with tasks requiring explicit instructions. To ensure consistency and repeatability across all evaluations, sampling was disabled in each model. Additionally, when multiple model variants were available (**Table 1**), the largest model was chosen, provided it did not exceed 30 billion parameters. For Mistral3.1, the longer edge of the input image was rescaled to a maximum of 1036 pixels to limit the number of image tokens and prevent excessive VRAM usage.

**Table 1 Overview of the evaluated vision-language models.**

| Model Name | Company | Country | Parameters in Billion | Modalities | Release Date |
|---|---|---|---|---|---|
| Phi-4 (Multimodal) | Microsoft | USA | 6 | Text, Image, Audio | February 2025 |
| Gemma3 | Alphabet (Google) | USA | 1; 4; 12; 27 | Text, Image | March 2025 |
| Qwen2.5 (VL) | Alibaba (Qwen) | China | 3; 7; 32; 72 | Text, Image | January 2025 |
| LLama3.2 (Vision) | Meta | USA | 11; 90 | Text, Image | September 2024 |
| Mistral3.1 (Small) | Mistral AI | France | 24 | Text, Image | March 2025 |

## Prompt Design

To evaluate diagnostic reasoning and multimodal performance, each model was tested under three prompting configurations: baseline, multimodal and chain-of-thought reasoning.

The baseline prompt was "Choose the correct answer and return the following JSON template: <schema>", where <schema> represents a JavaScript Object Notation (JSON)-formatted string listing all possible diagnostic labels for the task (e.g. for the Endo dataset: "{'tumor': ['Yes', 'No']}").

The multimodal prompting configuration was applied only to the neonatal jaundice dataset as it was the only dataset that included both relevant clinical context and images. Furthermore, the clinical context (serum bilirubin concentration) is an easily interpretable indicator of neonatal jaundice, enabling an evaluation of the model's ability to prioritize the most diagnostically relevant information. In this setting, clinical context was prepended to the baseline prompt in the form: "This is a patient with potential neonatal jaundice with the following clinical characteristics: <clinical>.

<baseline>", where <clinical> represented structured clinical data such as "serum bilirubin (mg/dL): 3.9".

The chain-of-thought reasoning prompt extended the baseline prompt by appending an explicit request for justification: "<baseline>. Explain your choices first." This modified prompt instructed the model to generate a rationale before providing its final answer.

## Statistical Analysis

Statistical analysis was performed by G.M.-F. and D.T., who have 5 and 17 years of relevant experience, respectively. All analyses were conducted using Python version 3.10 with the SciPy library version 1.15.

For each model within each experimental setting, model performance was assessed by comparing the model's output for each task against the ground truth labels provided in the datasets. Model predictions were categorized as True Positives (TP), True Negatives (TN), False Positives (FP), and False Negatives (FN) based on a binary classification framework. The diagnostic performance of each model in each setting was quantified using accuracy, calculated as the overall proportion of correct predictions, using the formula (TP+TN)/(TP+TN+FP+FN).

To statistically compare the performance of the models within each experimental setting, a bootstrapping approach was employed. This involved performing 1,000 resamplings of the dataset with replacement. For each resampling, the accuracy was calculated for each model. The difference in accuracy between any two models being compared was then determined for each resampling, generating a bootstrap distribution of the difference. A 95% confidence interval for this difference was constructed using the percentile method from the bootstrap distribution. A P-value of less than 0.05 was considered statistically significant for these comparisons.

## Results

### Visual Input Only

When only visual (image) input was used, Qwen2.5 achieved an accuracy of 90.4% [90.2%–90.6%] across all 19 subclasses of the ChestDR dataset and 84.2% [83.6%–84.7%] on the Endo dataset (**Table 2** and **Figure 1**). For both datasets, Qwen2.5 had significantly higher accuracy than all other models tested (all P < 0.001).

On the ColonPath dataset, the accuracy of Phi-4 (69.6% [68.6%–70.4%]) was not significantly different from that of Qwen2.5 (69.0% [68.1%–70.0%], P = 0.410), and both models significantly outperformed the other three (all P < 0.001). Similarly, for the NeoJaundice dataset, the accuracy of Qwen2.5 (58.3% [56.2%–60.2%]) was not significantly different from that of Phi-4 (58.1% [56.1%–60.0%], P = 0.934), and both were superior to the other models (all P < 0.001).

**Table 2: Model accuracy on five medical imaging datasets.**

|  | **Phi-4** | **Gemma3** | **Qwen2.5** | **Llama3.2** | **Mistral3.1** |
|---|---|---|---|---|---|
| ChestDR | 66.4% [66.1%-66.7%] | 75.0% [74.7%-75.2%] | **90.4% [90.2%-90.6%]** | 22.8% [22.6%-23.1%] | 89.9% [89.7%-90.1%] |
| ColonPath | **69.6% [68.6%-70.4%]** | 64.7% [63.7%-65.6%] | 69.0% [68.1%-69.9%] | 34.5% [33.5%-35.4%] | 29.9% [29.0%-30.8%] |
| Endo | 79.3% [78.8%-79.9%] | 53.5% [52.8%-54.3%] | **84.2% [83.6%-84.8%]** | 54.7% [53.9%-55.5%] | 66.0 [65.3%-66.7%] |
| NeoJaundice | 58.1% [56.1%-60.0%] | 30.1% [28.1%-31.9%] | **58.3% [56.3%-60.3%]** | 41.7% [39.5%-43.6%] | 30.9% [28.9%-32.8%] |
| Retino | 0.1% [0.0%-0.2%] | 18.6% [16.6%-20.6%] | **18.6% [16.7%-20.6%]** | 8.0% [6.6%-9.6%] | 9.4% [8.0%-11.1%] |

Values represent mean accuracy (%) with 95% confidence intervals. Results are based on image-only input. The highest accuracy per dataset is shown in bold; statistical significance is discussed in the text.

For the Retino dataset, no model was able to achieve an accuracy higher than 20%. Qwen2.5 achieved an accuracy of 18.6% [16.7%–20.7%], which was significantly higher (P<0.001) than that of all models except Gemma3, which matched its accuracy at 18.6% ([16.6%–20.6%], P = 0.994).

Class-wise performance analyses for the ChestDR and Endo datasets are provided in **Figures S1** and **S2** of the Supplementary Materials. Overall, the models exhibit broadly similar performance profiles across individual classes within each dataset, suggesting stable model rankings. However, a notable exception is observed for Gemma3, which shows a marked drop in accuracy for the class label *"increased lung markings"*, indicating a specific challenge for the model in predicting this class accurately.

The diagnostic performance of the 7B parameter Qwen2.5 model was compared with smaller (Qwen2.5-3B) and larger (Qwen2.5-32B) variants to evaluate the effect of model size on medical image understanding. The results suggest that model performance does not simply scale with size, as shown in **Figure S3** of the Supplementary Materials. The 3b variant was comparable or inferior to Qwen2.5, with statistically significant lower accuracy on ColonPath (p = .004), NeoJaundice (p = .01), and Retino (p < .001), while showing no significant difference on ChestDR (p = .93) and Endo (p = .76). Notably, the 32B model achieved significantly higher accuracy only on the Retino dataset (p < .001), but was otherwise comparable to Qwen2.5 on ChestDR (p = .10) or inferior on ColonPath, Endo, and NeoJaundice (all p < .001).

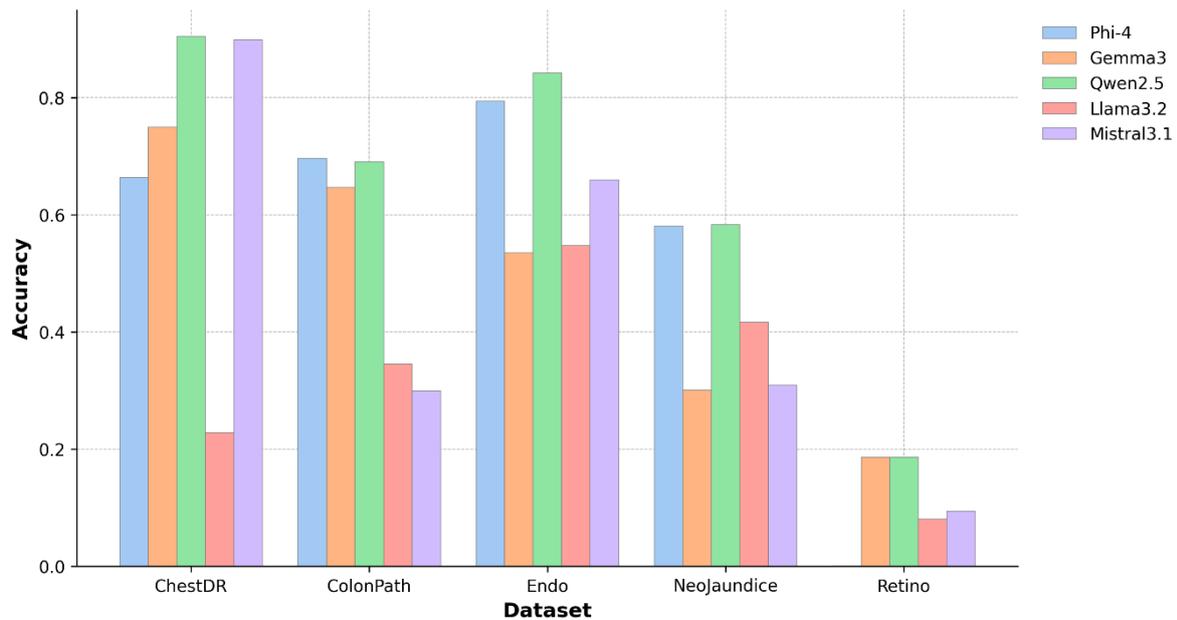

**Figure 1: Model accuracy across five medical imaging datasets.** Solid bars indicate the accuracy of each model. Qwen2.5 consistently achieve the highest accuracies in all datasets, while Llama3.2 shows generally lower performance across all datasets. The Retino dataset presents the greatest challenge, with all models performing notably worse compared to the other datasets.

## Multimodal Input

To evaluate the models' ability to handle different input modalities, each model was tested on the NeoJaundice dataset under three conditions: (1) image-only input (baseline), (2) clinical-only input using the reported bilirubin level (mg/dL), and (3) a multimodal configuration combining both image and clinical data. Compared to the baseline, the multimodal configuration led to a statistically significant improvement in accuracy for all models except Phi-4 ($p < .001$; see **Figure 2**). The accuracy of Phi-4 was not significantly different between the multimodal and baseline conditions (55.6% [53.6%-57.7%] vs. 58.1% [56.0%–60.4%], $p = .11$).

Interestingly, Qwen2.5, Llama3.2 and Gemma3 achieved higher accuracy with clinical-only input than with multimodal input ($p < .001$), suggesting that for these models, the addition of image data may have interfered with clinical reasoning. As expected, a bilirubin threshold of 12.9 mg/dL perfectly separated the patient classes in the dataset, demonstrating that Qwen2.5 correctly interpreted the clinical decision boundary by achieving almost 100% accuracy using clinical data alone.

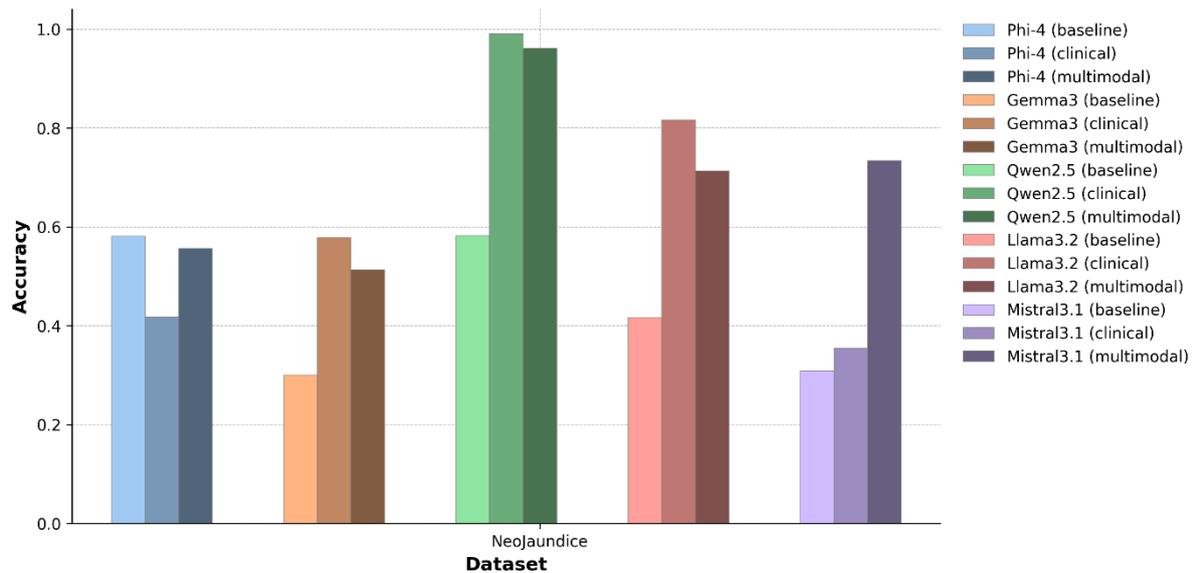

**Figure 2: Model accuracy on the NeoJaundice dataset as a function of input type.** The "baseline" configuration exclusively used visual input for diagnosis. The "clinical" configuration exclusively used the bilirubin level as input. The "multimodal" configuration used both visual and clinical information as inputs.

## Chain-of-thought Reasoning

When the models were prompted to first reason before answering, accuracy varied substantially across datasets, with most models performing worse than their baseline classification accuracies (**Figure 3**). Despite its strong baseline classification performance, Qwen2.5 showed significantly reduced accuracy in colon pathology, retinal fundoscopy, and endoscopy ($p < .001$) and no significant difference in neonatal jaundice when prompted to reason first. Reasoning improved the accuracy of Llama3.2 in colon pathology ($p < .001$) but had a negative or no effect on the accuracy of other tasks. Gemma3 also demonstrated mixed performance, with improved accuracy in endoscopy ($p < .001$) but significant drops in accuracy in other datasets. Phi-4 consistently failed to reason and return answers in valid JSON format, performing at or near zero across all datasets and significantly below its baseline in every case ($p < .001$). Mistral3.1 was the only model to consistently improve with reasoning, outperforming its baseline on three of five evaluated datasets (colon pathology, retinal fundoscopy, and endoscopy; all $p < .001$). **Figure 4** provides an example of this improvement: Mistral3.1 correctly identified a polyp only when prompted to explain its reasoning first. However, it also incorrectly identified the polyp as a tumor, demonstrating that while prompting for reasoning can enhance certain aspects of diagnostic accuracy, it may also introduce new classification errors.

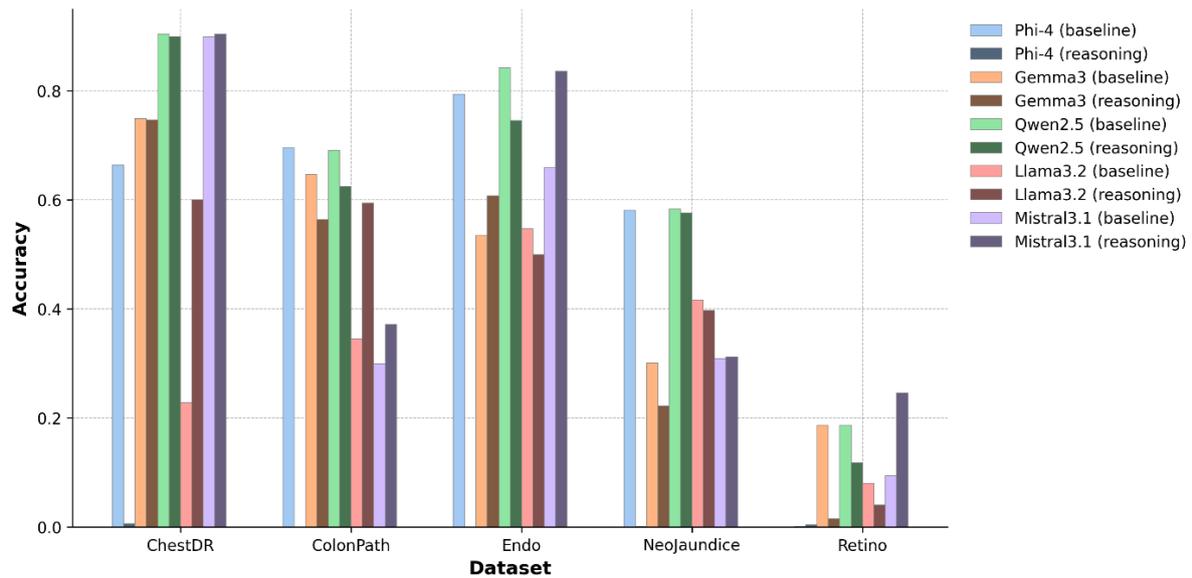

**Figure 3: Comparison of model accuracy under baseline and reasoning configurations across clinical imaging tasks.** The "baseline" configuration reflects performance using the standard prompt. The "reasoning" configuration uses a modified prompt that explicitly requests a step-by-step ("chain-of-thought") explanation prior to answer generation.

## Discussion

While proprietary, closed-source VLMs have shown advanced capabilities, their clinical application is often constrained by data privacy and governance concerns. The distilled versions evaluated here offer the possibility of local deployment within institutional boundaries, but these general-purpose models are not specifically pre-trained or fine-tuned on medical data. This comprehensive evaluation of state-of-the-art open-source vision-language models (VLMs) on medical image interpretation tasks reveals that performance varies across models and datasets. Although some models, e.g., Qwen2.5, demonstrated promising capabilities in specific imaging contexts, substantial challenges, particularly in the areas of complex image interpretation and effective multimodal data integration, remain to be addressed before they can be widely adopted in clinical practice.

In image-only tasks, the accuracy of Qwen2.5 was higher than or comparable to that of the other evaluated models. Importantly, Qwen2.5 adhered to structured output formatting (JSON), a crucial factor for seamless integration into clinical informatics systems. However, all models struggled with grading diabetic retinopathy fundoscopy images, with accuracies below 20%. This low accuracy indicates that general-purpose VLMs without specialized fine-tuning are not clinically applicable for such nuanced pathological features. These nuanced pathological features include microaneurysms, hemorrhages, exudates, and neovascularization that require expert-level interpretation(17).

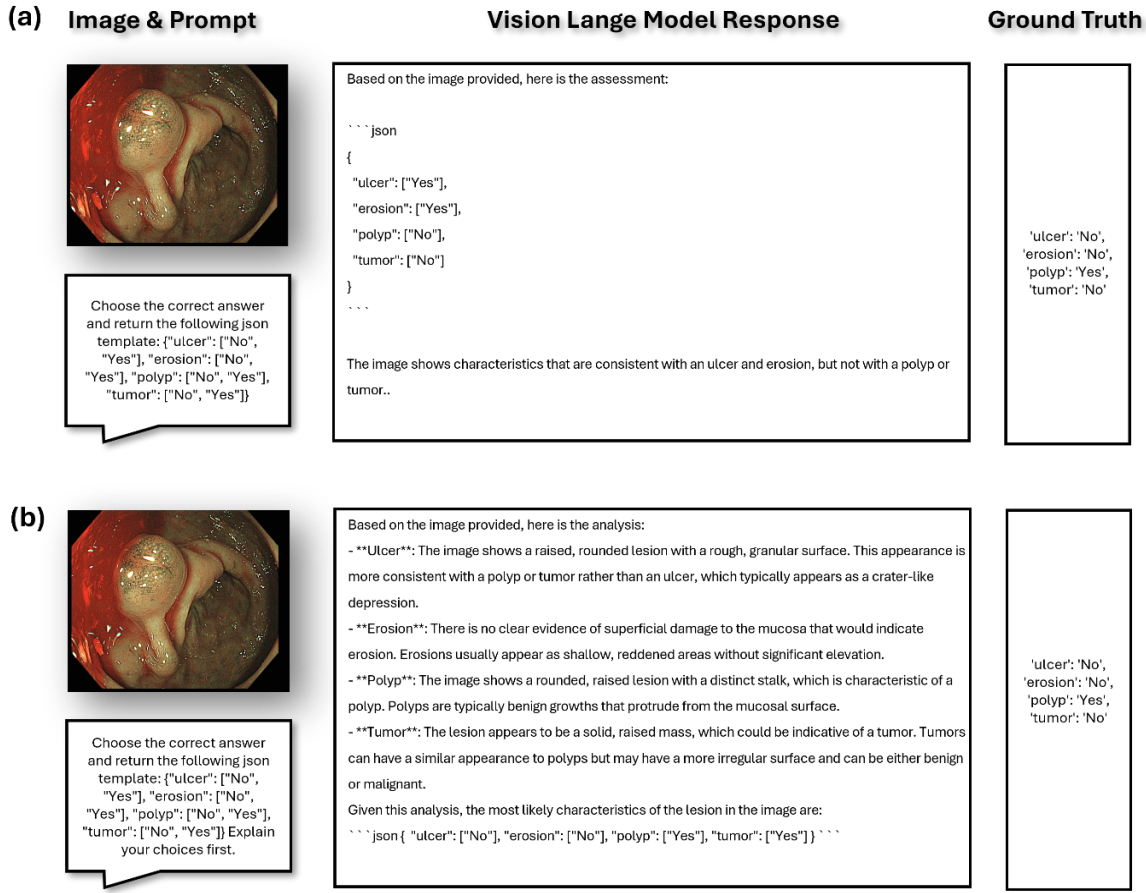

**Figure 4: Comparative assessment of a gastrointestinal lesion based on endoscopic imagery using Mistral3.1 without (a) and with instructions to provide reasoning (b).** In panel a, without prompting for explanation, Mistral3.1 misclassifies the lesion by incorrectly identifying both an ulcer and erosion while failing to detect the presence of a polyp. In panel b, when the model is instructed to explain its reasoning prior to classification, it correctly identifies the lesion as a polyp but incorrectly labels it as a tumor.

The investigation of multimodal data integration using the NeoJaundice dataset revealed that incorporating clinical parameters alongside imaging generally enhanced diagnostic accuracy for most models. This task—detecting jaundice based on serum bilirubin levels—is relatively simple; thus, using this dataset reveals whether the models can prioritize the most clinically relevant information when presented with both visual and non-visual inputs. A critical observation is that the accuracies of Qwen2.5 and Gemma3 were higher under clinical-only input than under multimodal input. This suggests that visual features may introduce noise or distract from decisive clinical indicators, such as bilirubin levels. This important finding demonstrates that imaging can, in some cases, confuse rather than aid diagnosis when laboratory values alone are sufficient. This phenomenon warrants further investigation, as effective fusion of multimodal data remains a critical objective for clinical VLMs(18,19).

Chain-of-thought reasoning is expected to improve diagnostic accuracy by guiding models through structured, step-by-step inference, allowing them to break down complex tasks into manageable and interpretable components(20,21). Contrary to our expectation that prompting for explicit reasoning would enhance diagnostic accuracy, most models, including Qwen2.5, exhibited a decline in accuracy across multiple datasets. Mistral3.1 was a notable exception, consistently improving its performance with reasoning prompts across three of the five evaluated datasets. This suggests that its training may be more amenable to such explicit reasoning steps. Conversely, Phi-4 consistently failed to provide valid outputs when prompted to explain its choices first, highlighting a strong limitation in its current instruct-following capabilities. Consistent with the results of recent studies(22,23), these findings indicate that additional fine-tuning for reasoning tasks is required to enhance the performance of open-source VLMs.

This study has several limitations. First, the evaluation was retrospective and based on a single publicly available dataset, which only partially captures the complexity, heterogeneity, and noise characteristic of real-world clinical data. Second, our study focused on diagnostic classification tasks, and performance on other clinically relevant tasks such as lesion localization, overall survival prediction, or automated report generation remains unevaluated. Finally, the selection of models was constrained by the parameter limits set in our methodology and did not cover all publicly available models.

In conclusion, recently released open-source VLMs demonstrate a range of capabilities for medical image interpretation, with Qwen2.5 showing particular promise in several tasks. However, the overall landscape indicates that "off-the-shelf" performance is not yet sufficient for reliable, unsupervised clinical use. Performance is highly variable and task-dependent, with significant challenges remaining in complex image interpretation, effective multimodal data integration, and reasoning. While the potential of open-source VLMs as locally deployable, privacy-preserving tools in healthcare is substantial, considerable development, domain-specific adaptation, and rigorous validation are necessary before these models can be safely and effectively integrated into routine radiological practice.

# Supplementary Materials

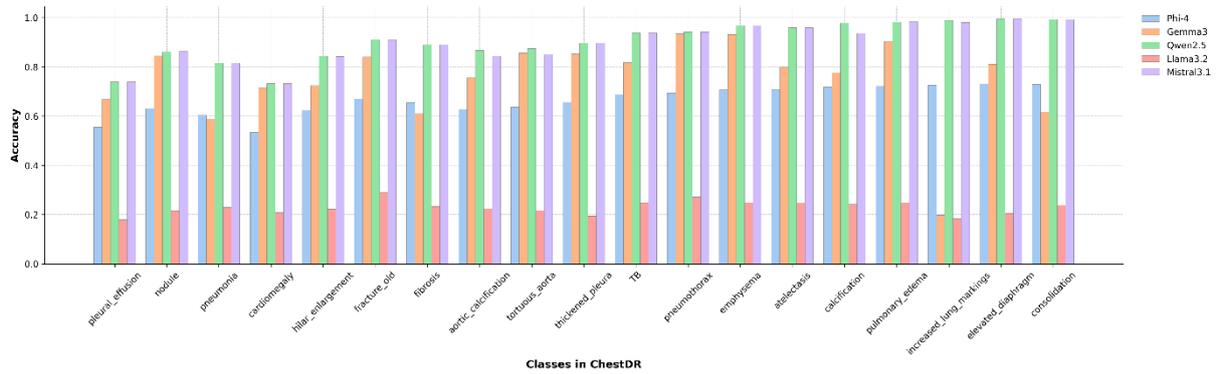

**Figure S1: Model accuracy in the ChestDR dataset separated by classes.** Solid bars indicate the accuracy of each model.

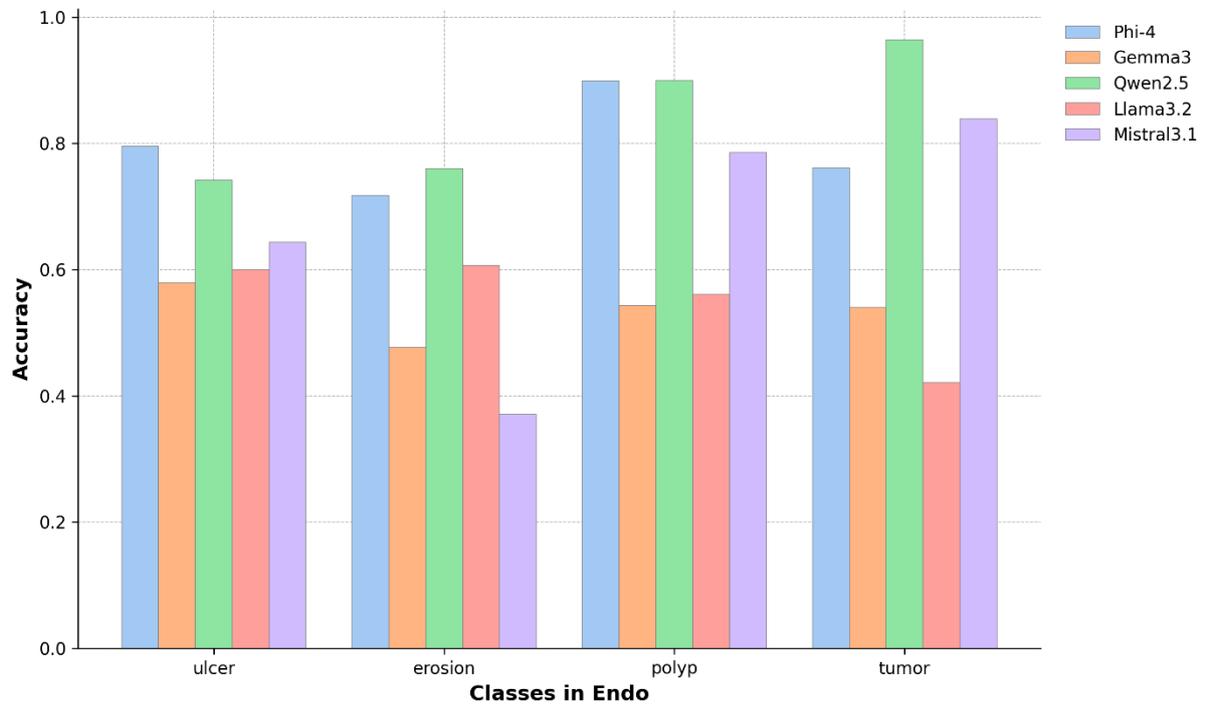

**Figure S2: Model accuracy in the Endo dataset separated by classes.** Legend as in Figure S1.

Model Size

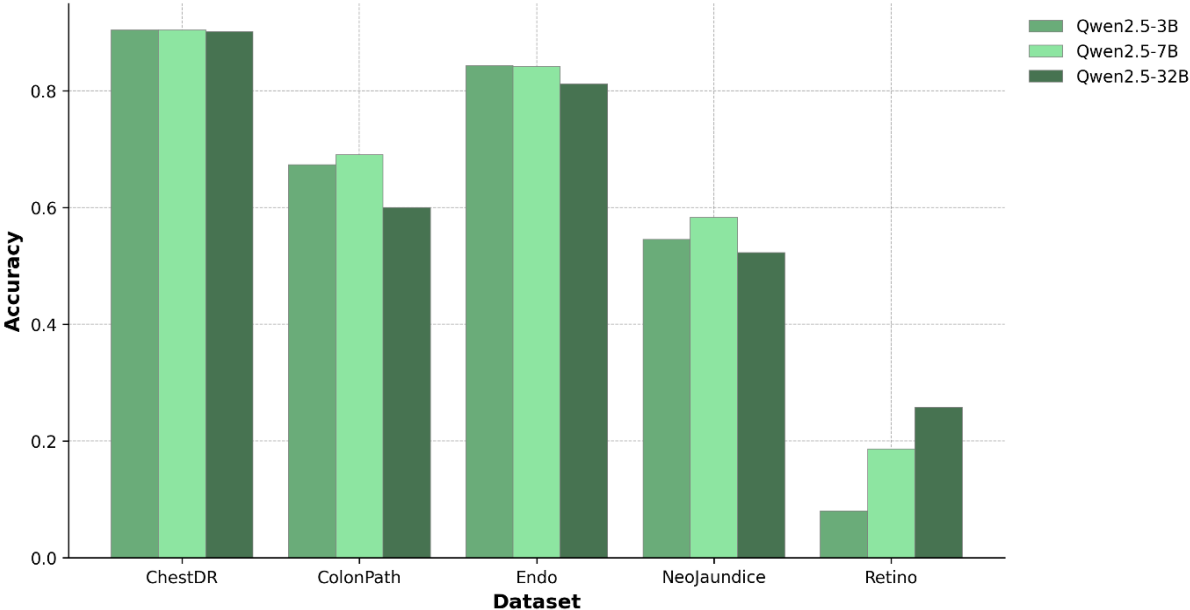

**Figure S3**: **Qwen2.5 accuracy across five medical imaging datasets as a function of model parameters.** The default configuration "Qwen2.5-7b" has 7 billion parameters, "Qwen2.5-3b" has 3 billion parameters and "Qwen2.5-32b" has 32 billion parameters.